\documentclass[preprint,aps,prl,showpacs,twocolumn,mamsmath,10pt]{revtex4}
\usepackage{graphics}
\usepackage{psfrag}
\usepackage{epsfig}
\usepackage{epsf}
\usepackage{float}
\usepackage{amssymb,stmaryrd,latexsym}

\newcommand{\be}{\begin{eqnarray}}
\newcommand{\ee}{\end{eqnarray}}
\usepackage{amsmath}
\usepackage{amssymb}
\usepackage{epsfig}

\begin{document}
\preprint{\small FZJ--IKP(TH)--2008--05, HISKP-TH-08/05}

\title{Evidence that the $\mathbf{ Y(4660)}$ is a {\boldmath$f_0(980)\psi'$} bound state}

\author{Feng-Kun Guo$^1$, Christoph~Hanhart$^1$, and Ulf-G. Mei\ss ner$^{1,2}$}

\affiliation{$^1$ Institut f\"{u}r Kernphysik, Forschungszentrum J\"{u}lich GmbH,
 52425 J\"{u}lich, Germany}
\affiliation{$^2$ Helmholtz-Institut f\"{u}r Strahlen- und Kernphysik,
Universit\"at Bonn,
53115 Bonn,
Germany }

\begin{abstract}
\noindent
We demonstrate that the experimental information currently
available for the Y(4660) is consistent with its being
a $f_0(980)\psi'$ molecule. Possible experimental tests
of our hypothesis are presented.
\end{abstract}

\maketitle

{\bf 1.} In recent years a large number of new hidden and open charm
states were discovered experimentally above the first inelastic
thresholds. For most of them the masses have in common that they deviate
significantly from the predictions of the quark
model~\cite{quarkmodel} --- which on the other hand works very well
below the thresholds --- and are positioned very close to a
$s$-channel threshold~\cite{threshold}.
 For recent reviews see, e.g., Ref.~\cite{charmreview}.

The proximity of the thresholds initiated a lot of theoretical
studies, in order to reveal a possible molecular structure of those
states. E.g. in Refs.~\cite{dsrefs} the $D_s(2317)$, located just
below the $KD$--threshold, was studied within the molecular model
and the $X(3872)$, located right at the $\bar DD^*$ threshold, was
investigated in  Refs.~\cite{X3872}. However, so far no consensus
exists on the true nature of those states and, e.g., four--quark
interpretations~\cite{tetraq} and even conventional $\bar qq$ states
are still under discussion~\cite{qqbar}.

In Refs.~\cite{moltheory} it was
argued that there is a way to model
independently identify hadronic molecules
in the spectrum. The method is based
on a time--honored analysis by Weinberg~\cite{wein} and
applies, if the pole is very close to
the threshold of the constituent particles
that form the bound state in an $s$--wave.
The method relates the effective coupling
constant of the bound state to its constituents, $g$,
directly to the molecular admixture of the
state. Especially, one may write
for a pure molecule
\begin{eqnarray}
\frac{g^2}{4\pi}=
\frac{(m_1+m_2)^{5/2}}{(m_1m_2)^{1/2}}\sqrt{32\epsilon} \ ,
\label{geff}
\end{eqnarray}
where $m_1$ and $m_2$ denote the masses of the constituents
and $\epsilon$ the binding energy.

\begin{figure}[t]
\begin{center}
\psfrag{psi}{$\psi'(3686)$}
\psfrag{g}{$g$}
\psfrag{f}{$f_0(980)$}
\psfrag{Y}{$Y(4660)$}
\psfrag{p1}{$\pi^+/K^+/\gamma$}
\psfrag{p2}{$\pi^-/K^-/\gamma$}
\epsfig{file=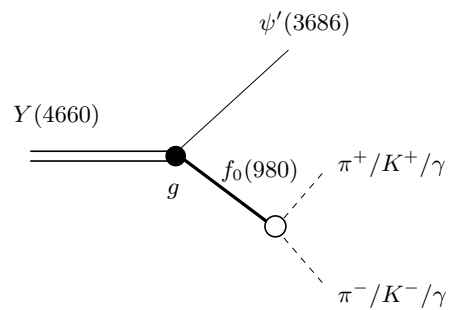, width=5cm}
\caption{Diagram illustrating the
most prominent decay channels of the $Y(4660)$
in the molecular model. The solid black dot
denotes the $Yf_0\psi'$ vertex, whose strength
parameter $g$
is fixed within the molecular model through Eq.~(\ref{geff}).
The decay vertices for $f_0$, shown as open circle,
are fixed from other data.\label{diag}
}
\end{center}
\end{figure}

In this paper we discuss the nature of the $Y(4660)$ as a candidate
for a $f_0(980)\psi'$ bound state. So far the $Y(4660)$ was
seen only in the $\pi^+\pi^-\psi'$ invariant mass distribution in
$e^+e^-\to\gamma_{ISR}\pi^+\pi^-\psi'$ with a mass of
$4664\pm11\pm5$~MeV and a width of $48\pm15\pm3$~MeV \cite{Yexp}. Such
a structure was neither observed in
$e^+e^-\to\gamma_{ISR}\pi^+\pi^-J/\psi$ \cite{ISRJpsi}, nor in the
exclusive $e^+e^-\to D{\bar D},D{\bar D}^*, D^*{\bar D}^*, D{\bar
D}\pi$ cross sections \cite{Exp:DDbar}, nor in the process $e^+
e^-\to J/\psi D^{(*)} {\bar D}^{(*)}$ \cite{Abe:2007sy}. These facts would severely
challenge any attempt explaining the state as a canonical $c{\bar
c}$ charmonium, e.g. $5^3S_1$ as in Ref. \cite{Ding}. The $Y(4660)$ is
suggested to be a baryonium state in Ref. \cite{Qiao}. The difficulties in
the canonical charmonium interpretation could be explained naturally,
if the $Y(4660)$ were a $f_0(980)\psi'$ bound state, because it would
decay dominantly via the decay of the $f_0(980)$, i.e.
$Y(4660)\to\psi'f_0(980)\to \psi'[\pi\pi]$ and
$Y(4660)\to\psi'f_0(980)\to \psi'[K{\bar K}]$. The nominal threshold of the
$f_0(980)$--$\psi'$ system is at $4666\pm10$~MeV if we take the
PDG value of the $f_0(980)$ mass \cite{PDG}. Note that the
interaction between a heavy quarkonium and a light hadron is
expected to be dominated by the QCD van der Waals interaction which
is attractive \cite{Brodsky}.

In the $\pi\pi$ invariant mass a clear $f_0(980)$ peak is visible.
Clearly, if the $Y(4660)$ were a conventional $f_0\psi'$ bound
state, it could not decay to this channel.
This is possible only because of the finite width of the $f_0(980)$,
 mainly due to its
decay to the $\pi\pi$ channel. We will perform our analysis based on
the following reasoning: if the $f_0$ were a stable particle, also
the $Y(4660)$ were stable. Then we could calculate the effective
coupling constant of $Y$ to $\psi' f_0$ using Eq.~(\ref{geff}). The
central assumption is that this coupling constant does not change as
the two pion channel is switched on
--- a similar ansatz lead to a successful phenomenology
for the $f_0(980)$ treated as a $\bar KK$ molecule~\cite{f0}.
 Under this assumption we can
predict not only the prominent component of the width of the
$Y(4660)$ but also its spectral shape using the mass and the
over all normalization as the only
input. As additional non--trivial result we can
also predict the strength and the shape of the decay to $\bar KK$
and $\gamma\gamma$. The possible decay chains are illustrated in Fig.~\ref{diag}.
As we will see, the resulting spectra are in excellent
agreement with the existing data. We interpret this as
a strong support in favor of a molecular interpretation
for the $Y(4660)$.

{\bf 2.} The $Y(4660)$ was observed as a structure
in the $\psi'\pi\pi$ invariant mass distribution.
In addition, in Ref.~\cite{Yexp}, also the $\pi\pi$
invariant mass distribution is presented, which
was found after applying an appropriate cut
to the $\psi'\pi\pi$ invariant mass.
Both distributions can be derived from
the same differential rate,
 under
the assumption that the pion pair stems from
the decay of a $f_0$. One gets
\begin{eqnarray}
\frac{d^2{\mathcal W}(e^+e^-{\to}\psi'\pi^+\pi^-)}{dM^2dm_{\pi\pi}^2}
{=} N\left|G_Y(M)\right|^2\frac{d\Gamma_Y^{[\pi^+\pi^-]}}{dm_{\pi\pi}^2} \, ,
\label{rate}
\end{eqnarray}
where $G_Y(M)$ denotes the physical $Y$--propagator, to be specified
below and
 $N$ is a normalization constant, which contains, besides the electron--photon
vertex and the photon propagator, both indeed constant to very high accuracy in the
range of $M$ of relevance here, also the detector acceptance.
The decay $Y\to \psi'\pi^+\pi^-$ is described by
\begin{equation}
\frac{d\Gamma_Y^{[\pi^+\pi^-]}}{dm_{\pi\pi}^2}
 = \frac{g^2}{8\pi}
\theta(M-M_{\psi'\pi\pi})
\frac{p}{M^2}\rho_{f_0}^{[\pi^+\pi^-]} \ .
\label{width}
\end{equation}
Here $p$ 
 denotes the c.m.s momentum of the $\psi'$
 for given values of the $\pi\pi$ invariant mass, $m_{\pi\pi}$,
and the $\psi'\pi\pi$ invariant mass, $M$,
and $M_{\psi'\pi\pi}=m_{\pi\pi}+M_{\psi'}$.
The effective coupling constant $g$ is not a free parameter,
but can be determined for any given value of the mass of the
$Y$ from Eq.~(\ref{geff}).
The quantity $\rho_{f_0}^{[\pi^+\pi^-]}(m_{\pi\pi})$
denotes the $\pi^+\pi^-$ fraction of the $f_0$ spectral function.
It is normalized according to
$$
\int dm_{\pi\pi}^2 \rho_{f_0}^{[\pi^+\pi^-]}(m_{\pi\pi}) =
\Gamma_{f_0}^{[\pi^+\pi^-]}/\Gamma_{f_0}^{\rm tot} \ .
$$
With this normalization Eq.~(\ref{width}) goes to the
standard expression for a two particle decay in the
stable particle limit for the $f_0$.

A high quality data set for the $f_0$ was
collected recently by KLOE~\cite{kloe} based on the reaction
$\phi\to \gamma \pi\pi$. The data was analysed using
the so-called kaon loop model~\cite{achasov} and
provided parameters for the $f_0$ with very little
uncertainty. To be concrete, we use
\begin{equation}
\rho_{f_0}^{[\pi^+\pi^-]}(m) = \frac1{\pi}\frac{{\rm
Im}(\Pi^{\pi^+\pi^-}_{f_0}(m))} {\left| m^2-m_{f_0}^2+\sum_{ab}\hat
\Pi^{ab}_{f_0}(m)\right|^2} \ ,
\end{equation}
where the $\hat \Pi^{ab}_{f_0}(m)=\Pi^{ab}_{f_0}(m) -{\rm
  Re}(\Pi^{ab}_{f_0}(m_{f_0}))$ denote the renormalized self--energies
of the $f_0$ with respect to the channel $ab$. Analytic expressions
are given in Ref.~\cite{achasov}(see also 
\footnote{We only include the $\pi\pi$
  and the $\bar KK$ channels, for the others give a negligible
  contribution.}). The input parameters are taken from the fits
provided in Ref.~\cite{kloe}.  To be concrete, for all the curves
shown below we used the central values of the various parameters of
fit $K2$ shown in table 4 of that reference, thus we used
$m_{f_0}=0.9862$ GeV, $g_{f_0K^+K^-}=3.87$~GeV, and
$g_{f_0\pi^+\pi^-}=-2.03$ GeV, while the couplings for the
neutral channels were fixed using the isospin relations.
We checked that the other parameter sets give very similar results
to the ones discussed in detail below.

The only missing piece is the physical $Y$ propagator. In
Eq.~(\ref{width}) an explicit expression is given for the partial
width to $\psi'\pi^+\pi^-$. However, in order to derive the
physical propagator, all relevant decay channels need to be
included. Within our model we assume that the $Y$ decays
predominantly through the $f_0$, thus a consistent treatment
requires that all decay channels of the $f_0$ also contribute to the
width of the $Y$ and thus to the propagator. It is straightforward
to extend Eq.~(\ref{width}) also to the $\pi^0\pi^0$ channel as well
as to the kaon channels without any additional free parameters ---
also for these channels we use the results of the fit to the KLOE
data presented in Ref.~\cite{kloe}. Thus we get for the total width
of the $Y$ at a given value of $M$, under the assumption that it is
saturated by the $\psi'f_0$ decay
\begin{eqnarray}
\nonumber
\Gamma^{\rm tot}_Y(M^2)&{=}&\sum_{ab}
\theta\left(M-M_{\psi'}-\sqrt{s_{\rm thr}^{ab}}\right) \\
&{\times}&
\int_{s_{\rm thr}^{ab}}^{(M-M_{\psi'})^2} \!\!\!\!\! dm_{ab}^2
\frac{d\Gamma_Y^{[ab]}(M,m_{ab})}{dm_{ab}^2} \ ,
\end{eqnarray}
where $s_{\rm thr}^{ab}=(m_a+m_b)^2$.
In order to get a $Y$ propagator with the correct
analytical properties we need to continue the contribution
from the
$\bar KK\psi'$ channel also to below its threshold.
For this we use a dispersion
integral, which gives us an expression for the $Y$ self--energy, $\Pi_Y(M)$,
 for arbitrary
values of $M$
\begin{equation}
\Pi_Y(M) = \frac{1}{\pi}\int_{M_{\rm thr}^2}^\infty \!\!\!\!
ds\frac{M_Y\Gamma^{\rm tot}_Y(s)}{s-M^2-i\epsilon} \ ,
\end{equation}
where $M_{\rm thr}=M_{\psi'}+2m_\pi$ denotes the lowest physical threshold of
relevance here. Note, this treatment is completely consistent to what
was done for the $f_0$ --- one way to derive the self-energies $\Pi^{ab}_{f_0}(m)$
given above is through a dispersion integral with the two--body phase
space as input.
With the self energies at hand we may now give the expression for
the physical propagator of the $Y(4660)$
\begin{equation}
G_Y(M)=\frac{1}{M^2-M_Y^2+\hat \Pi_Y(M)} \ ,
\label{yprop}
\end{equation}
where, as above, we defined $\hat \Pi_Y(M) = \Pi_Y(M)-{\rm Re}(\Pi_Y(M_Y))$.

\begin{figure}[t]
\begin{center}
\psfrag{psi}{$\psi'$} \psfrag{Y}{$Y(4660)$} \psfrag{p}{$\pi (K)$}
\epsfig{file=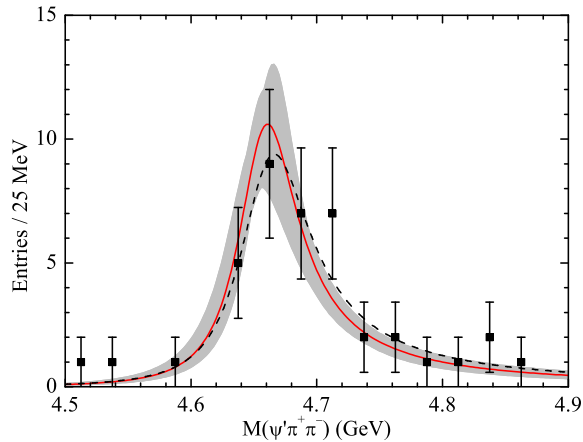, width=0.5\textwidth} \caption{Comparison of
the line shape in the $\psi'\pi^+\pi^-$ invariant mass distribution
derived from the molecular model with the data. The solid line shows
the result of our best fit, while the shaded area shows the
uncertainty that emerges from the fit. The dashed line shows the
best fit result, if also the effective coupling is part of the fit,
as discussed in the final section. \label{invmassspec} }
\end{center}
\end{figure}

{\bf 3.} Our model has only 2 free parameters,
namely, the mass of the $Y$, $M_Y$,
and the normalization constant, $N$, introduced in Eq.~(\ref{rate}).
We now proceed as follows: the count rate, $R$,
 in the $\psi'\pi^+\pi^-$ invariant mass
spectrum is given by
\begin{equation}
R(M){=}\int_{4m_\pi^2}^{M-M_{\psi'}}
\!\!\!\!\!\!\!\!
dm_{\pi\pi}^2
\frac{d^2{\mathcal W}(e^+e^-\to\psi'\pi^+\pi^-)}{dM^2dm_{\pi\pi}^2} \ ,
\end{equation}
using the expression for the differential rate ${\mathcal W}$ of Eq.~(\ref{rate}).
The range of parameters allowed by the
data is then determined from a fit to the experimental data for the $\psi'\pi^+\pi^-$
spectrum --- we remind
the reader that the coupling constant $g$ is fixed from Eq.~(\ref{geff})
for any given value of $M_Y$.
We find
\begin{eqnarray}
N= 10\pm 2 \,  {\rm GeV^3} \, , \  M_Y=4665^{+3}_{-5} \, {\rm MeV}.
\end{eqnarray}
This range of mass values corresponds to a range of $g=11 \ldots 14$
GeV for the effective coupling constant. The best fit is shown as
the solid line in Fig.~\ref{invmassspec} --- the uncertainty that
emerges from the fit is reflected by the grey band. For the best fit
$\chi^2/{\rm d.o.f} = 0.5$. The first observation is that the
resulting invariant mass distribution visibly deviates from a
standard Breit--Wigner shape. This is a direct consequence of our
starting assumption, namely that the $Y(4660)$ is predominantly
composed of a $f_0$ and a $\psi'$, not only since the mass of the
$Y$ is very close to the nominal $f_0\psi'$ threshold, but also
because of the proximity of the kaon channels, which are very
important for the structure of the $f_0$.

\begin{figure}[t]
\begin{center}
\epsfig{file=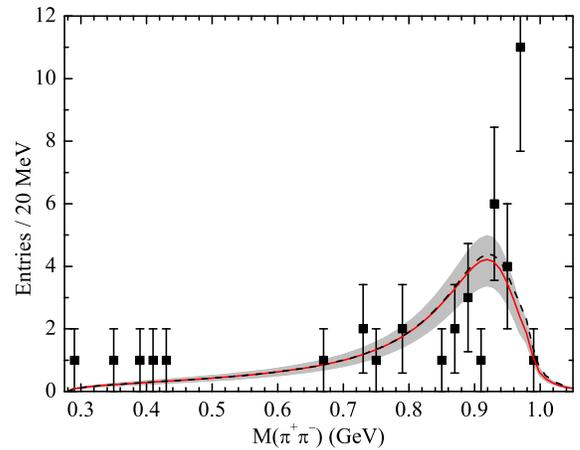, width=.5\textwidth} \vglue-1.2cm
\epsfig{file=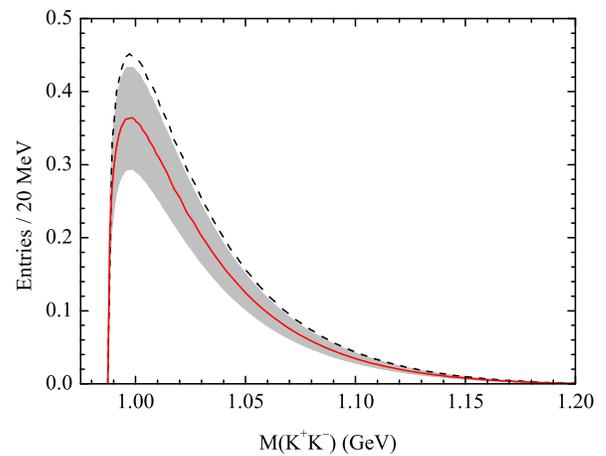, width=.5\textwidth} \caption{Predictions
for the $\pi\pi$ and $\bar KK$ invariant mass distributions.
\label{pipiKK} }
\end{center}
\end{figure}

 Since now
all parameters of our model are fixed we can predict other channels.
Recently high quality data was measured for $f_0\to
\gamma\gamma$~\cite{bellef0gg}. Within our approach we thus predict
that the shape in the $\psi'\gamma\gamma$ invariant mass is
identical to that measured for $\psi'\pi^+\pi^-$, however, down
scaled by the relevant ratio of branching ratios. In addition we can
also predict the signals for the spectra of $\pi\pi$ and $K\bar K$
as they emerge from the decay of the $Y$. The predicted rate,
$R_{ab}(m_{ab})$, again follows directly from Eq.~(\ref{rate})
\begin{equation}
R_{ab}(m_{ab}){=}\int_{(4.5 \ {\rm GeV})^2}^{(4.9 \ {\rm GeV})^2}
\!\!\!\!\!\!\!
dM^2
\frac{d^2{\mathcal W}(e^+e^-\to\psi'ab)}{dM^2dm_{ab}^2} \ ,
\end{equation}
where the limits of integration are chosen identical to the cuts
used to get the experimental rate~\cite{Yexp}.
 The corresponding results
are shown in Fig.~(\ref{pipiKK}).
If a signal of the given shape and strength were found in the
$\psi'\bar KK$ invariant mass distribution, it would provide
a strong support for the assumed prominent role of $f_0\psi'$
for the structure of the $Y$.

{\bf 4.} To summarize, we calculated the invariant mass spectrum for
$\psi'\pi^+\pi^-$ as well as the corresponding $\pi\pi$ and $\bar
KK$ spectra in the mass range of the $Y(4660)$ under the assumption
that the $Y(4660)$ is a $f_0(980)\psi'$ bound state and
$f_0(980)\psi'$
 is its
only decay channel of relevance.
A very good description of both spectra, where data exist, was
achieved. Especially, we find a visible deviation from a
Breit-Wigner shape for the $\psi'\pi\pi$ spectrum, consistent
with the data, although data with better statistics
would be very welcome to strengthen this point.

In principle our method also allows one to estimate the possible
contribution of channels other than $f_0\psi'$ to the width
of the $Y(4660)$. One just needs to add the term $iM_Y\Gamma_{\rm add}$
to the denominator of $G_Y(M)$, defined in Eq.~(\ref{yprop}) and
repeat the fitting procedure. With this we get
$\Gamma_{\rm add}=(30\pm 30)$~MeV --- thus, before better
data is available no reliable bound for the possible
additional width can be deduced. However, it is important
to observe that the value extracted from the current data
is consistent with zero within the uncertainty.

We also checked what happens, if we do not fix the effective
coupling $g$ according to Eq.~(\ref{geff}), but allow it to float as
well. Then we find $M_Y=(4672\pm 9)$~MeV and $g=(13\pm 2)$~GeV with
$\chi^2/{\rm d.o.f} = 0.4$. The result of the best fit is shown as the
dashed line in Figs.~\ref{invmassspec} and \ref{pipiKK}.
Thus the effective coupling constant extracted from this three
parameter agrees to that found before, only that the three
parameter fit prefers a larger value of the mass, partially
inconsistent with a molecular picture
 --- for masses
at the higher end the mass of the $Y$ is even larger than
$M_{\psi'}+m_{f_0}$.
  Also here we need to conclude that more data
are needed to draw a more firm conclusion --- especially we
showed that the $\bar KK$ invariant mass distribution is
quite sensitive to the mass of the $Y(4660)$ --- c.f. lower panel
of Fig. \ref{pipiKK}.
It is in any case
important to stress that the fit calls for a large coupling constant
for $Y \to f_0\psi'$,
which is naturally explained by the assumption that the $Y(4660)$
is generated dynamically in the $f_0\psi'$ channel.

We take our results
as a strong evidence for a molecular  interpretation of the $Y(4660)$.
We stress that a measurement of the $\psi'\bar KK$ channels
as well as an improvement of the resolution of the existing data
would allow for non--trivial tests of our hypothesis.

\medskip

We thank N.N.~Nikolaev and A.~Nogga for useful discussions.  We are
very grateful to C.-Z. Yuan for providing us with the experimental
data.  This work is partially supported by the Helmholtz Association
through funds provided to the virtual institute ``Spin and strong
QCD'' (VH-VI-231) and by the EU Integrated Infrastructure Initiative
Hadron Physics Project under contract number RII3-CT-2004-506078.

\end{document}